\documentclass[prl,amsmath,superscriptaddress,twocolumn]{revtex4}

\pdfoutput=1 
\usepackage{epsfig}
\usepackage{hyperref}
\usepackage{amssymb}
\usepackage{amsmath}
\usepackage{amsfonts}
\usepackage{bm}
\usepackage{braket}
\usepackage{graphicx}
\usepackage{xcolor}

\newcommand{\nn}{\nonumber}
\newcommand{\ud}{{\textrm{d}}}

\begin{document}
\title{Quantum quenches in isolated quantum glasses out of equilibrium}

\author{S. J. Thomson}
\email{steven.thomson@polytechnique.edu}
\affiliation{Centre de Physique Th\'{e}orique, CNRS, Institut Polytechnique de Paris, Route de Saclay, F-91128 Palaiseau, France}
\affiliation{Institut de Physique Th\'{e}orique, Universit\'{e} Paris Saclay, CNRS, CEA, F-91191 Gif-sur-Yvette, France}
\author{P. Urbani}
\email{pierfrancesco.urbani@ipht.fr}
\affiliation{Institut de Physique Th\'{e}orique, Universit\'{e} Paris Saclay, CNRS, CEA, F-91191 Gif-sur-Yvette, France}
\author{M. Schir\'o}
\email{marco.schiro@ipht.fr}
\thanks{ On Leave from: Institut de Physique Th\'{e}orique, Universit\'{e} Paris Saclay, CNRS, CEA, F-91191 Gif-sur-Yvette, France}
\affiliation{JEIP, USR 3573 CNRS, Coll\`ege de France,   PSL  University, 11,  place  Marcelin  Berthelot,75231 Paris Cedex 05, France}
\date{\today} 

\begin{abstract}
In this work, we address the question of how a closed quantum system thermalises in the presence of a random external potential. By investigating the quench dynamics of the isolated quantum spherical $p$-spin model, a paradigmatic model of a mean-field glass, we aim to shed new light on this complex problem. Employing a closed-time Schwinger-Keldysh path integral formalism, we first initialise the system in a random, infinite-temperature configuration and allow it to equilibrate in contact with a thermal bath before switching off the bath and performing a quench. We find evidence that increasing the strength of either the interactions or the quantum fluctuations can act to lower the effective temperature of the isolated system and stabilise glassy behaviour.
\end{abstract}

\maketitle

\emph{Introduction - } 
Understanding how and why many-body systems can fail to reach thermal equilibrium is both of fundamental value, as it allows us to test the hypothesis underlying equilibrium statistical physics, and of practical interest. In fact systems which fail to equilibrate can often exhibit rich new dynamical phenomena not seen in typical thermal states~\cite{Weiss_nature06,GringScience12,Fausti_Science11,MitranoNature16} 

Two main mechanisms of ergodicity breaking in many-body quantum systems have emerged recently. On the one hand, quantum integrable systems have an extensive number of conserved charges and so do not thermalize to a state whose macroscopic properties are determined by only a few quantities (such as energy and density) \cite{EF16}. 
On the other hand, the interplay of disorder and interactions can given rise to a robust mechanism for ergodicity breaking,  the many-body equivalent of Anderson localisation, \emph{a.k.a.} Many Body Localization (MBL), which does not require fine tuning to (typically isolated) integrable points. The absence of thermalisation in MBL is related to an \emph{emergent} integrability~\cite{SerbynPapicAbaninPRL13_2,HNO14,RMS15, Mo16,ThomsonSchiroPRB18}.

In between those two limits, for which thermalization fails on all time scales, there is a huge class of systems for which thermalization is possible but only on very long timescales. These are glassy systems, whose dynamics display ergodicity breaking due to metastability. 
In this case, the dynamical evolution is trapped by exponentially many metastable states that forbid equilibration on short timescales.
In finite dimensions, such metastable states have a finite (but very long) lifetime, while in the mean field limit their lifetime diverges with the system size (or dimension) due to the divergence of the free energy barriers between them. 
Nevertheless those systems are never completely out of equilibrium since in the end they relax on timescales that scale exponentially in either the system size or dimension \cite{CC05, Ca09}.

In contrast with MBL and integrable systems, glassy systems do not depend crucially on isolation from their environment and indeed most investigations on the dynamical behavior of quantum glasses has focused on a dissipative setting, where the system is coupled to a thermal bath. Here  important progress has been achieved through the solution of simplified fully connected models \cite{Cugliandolo+99,Cugliandolo+01,Cugliandolo+02,Biroli+02,Cugliandolo+04,MarklandEtAlNatPhys11}. An interesting question which has received far less attention concerns the dynamics of isolated quantum glasses. Recently the properties of highly excited eigenstates of paradigmatic mean field models of quantum glasses and their resulting dynamics have been investigated numerically through exact diagonalization of finite size systems~\cite{Laumann+14}, analytically using forward scattering approximations~\cite{Baldwin+16,Baldwin+17,BaldwinEtAlPRB18}, and more recently through a mapping to Rosenzweig-Porter random matrix model~\cite{FaoroEtAl_AnnPhys19}. 
Yet, in the thermodynamic limit the dynamical behavior of those quantum mean field models can be solved exactly using field theory techniques similar to those well developed for classical models ~\cite{CC05}.

In this work we extend those techniques to the quantum case, by focusing on the unitary dynamics of the isolated spherical quantum $p$-spin model, a paradigmatic example of a mean-field glass, whose Hamiltonian 
\begin{align}\label{eqn:H}
\mathcal{H} &= \frac{1}{2m}\sum_i\Pi^2_i-{\cal J}(t)\sum_{i_1 < ... < i_p}^{N} J_{i_1...i_p} \sigma_{i_1}...\sigma_{i_p}
\end{align}
describes a set of spins $\sigma_{i}$ all-to-all coupled by random $p$-body interactions $J_{i_1...i_p}$ drawn from a Gaussian distribution with zero mean and unit variance. 
To make the model more tractable but still non-trivial, we treat the spins as continuous variables~\cite{Kosterlitz+76} and enforce the spherical constraint $\sum_{i}^{N} \sigma^{2}_i = N$ 
by adding a Lagrange multiplier (hereafter denoted $z$). We further add a conjugate momentum $\Pi_i$ where $\left[\Pi_i,\sigma_j\right]=i\hbar(t) \delta_{ij}$ are canonical commutation relations, and we allow $\hbar(t)$ to be time-dependent in order to be able to change the strength of quantum fluctuations - for details, see the Supplementary Material \cite{SM}.  
This model has been extensively studied in both its classical~\cite{Derrida80,Derrida81,Gross+84,Kirkpatrick+87a,Kirkpatrick+87b,Crisanti+93,Cugliandolo+93,Castellani+05} and quantum version, when coupled to a thermal bath \cite{Cugliandolo+98,Cugliandolo+99,Cugliandolo+01,Biroli+01,Cugliandolo+02,Cugliandolo+04,Cugliandolo+04b}. 
At low temperature it displays a \emph{dynamical} glass transition $T_d$ due to the emergence of long-lived glassy states. Below this temperature equilibration is never reached and the system ages forever (but not on exponential timescales).
The dynamical temperature is a decreasing function of the strength of quantum fluctuations, as one may expect \cite{Cugliandolo+01}.
Though the isolated dynamics of the quantum $p$-spin model have not previously been studied, the classical isolated dynamics was recently investigated  in~\cite{Cugliandolo+17}. 
\begin{figure}[t!]
\begin{center}
\includegraphics[width= \linewidth]{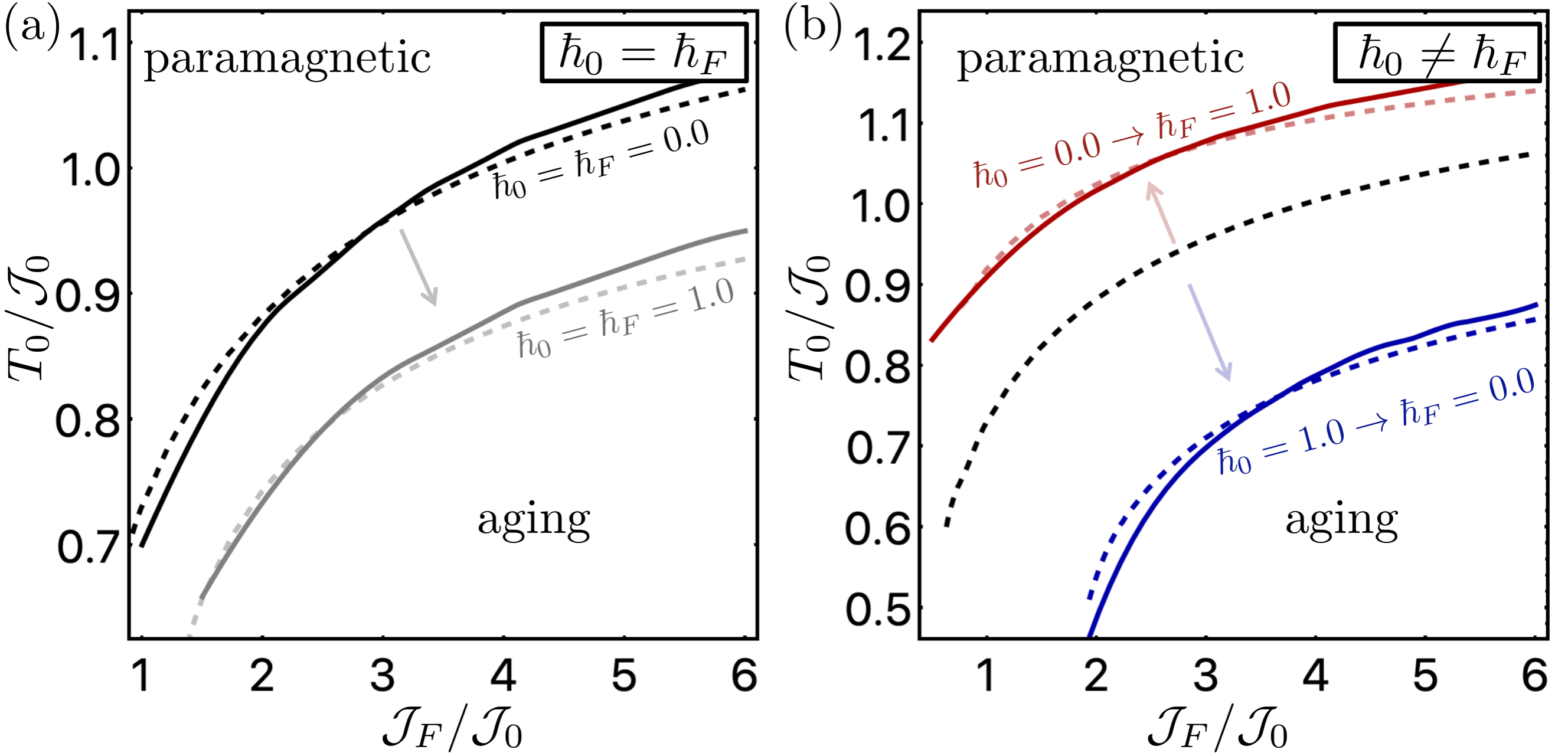}
\caption{Dynamical phase diagrams, as a function of initial temperature $T_0$ and strength of the interaction quench $\mathcal{J}_F/\mathcal{J}_0$, in two different scenarios. a) (Left panel) The strength of quantum fluctuations is kept constant throughout the evolution, i.e. $\hbar_0=\hbar_F$. In the classical case (top line) below a certain temperature the dynamics of the system displays  aging. Finite quantum fluctuations suppress the aging regime (bottom line), as expected thermodynamically. b) (Right panel)  When the strength of quantum fluctuations is suddenly increased,  $\hbar_F>\hbar_0$, the aging  regime is enhanced (top curve) with respect to the classical phase diagram (dashed line). Vice versa, decreasing quantum fluctuations makes the aging regime shrink (bottom curve).Details of how the boundaries were obtained are given in the main text.}
\label{fig.phase}
\end{center}
\end{figure}
Here we study the \emph{quantum} evolution of this model: we prepare the system at some temperature $T_0$ in the paramagnetic phase and then we suddenly change both the strength of random couplings $\mathcal{J}(t)$ and the strength of quantum fluctuations measured by $\hbar(t)$, keeping the system isolated. 
The resulting non-equilibrium phase diagram, plotted in Figure~\ref{fig.phase}, features a high-temperature paramagnetic phase, where the system relaxes toward equilibrium, and a low-temperature phase  where aging and breakdown of time-translational invariance emerge. Surprisingly, we find that the phase boundary between the paramagnetic and aging regimes strongly depends on whether quantum fluctuations are kept constant (left panel) or suddenly changed (right panel) throughout the evolution. In the former case the aging regime shrinks with respect to its classical counterpart, as expected thermodynamically. In the latter, we find that a sudden increase of quantum fluctuations promotes rather suppresses glassy effects (right panel, top curve), in striking contrast with the expectation based on the canonical equilibrium case of a system in contact with a finite temperature bath \cite{Cugliandolo+99,Cugliandolo+01,Cugliandolo+02}. Such enhancement of aging effects are due to an interplay of quantum fluctuations and non-equilibrium effects. We interpret this intriguing result in terms of an effective temperature $T_{eff}< T_{0}$ for the isolated disordered quantum system, which in the absence of an external thermal bath is able to cool itself down through quantum fluctuations,  eventually crossing the glass transition.

\emph{Dynamical Equations for Correlation and Response - } 
Throughout this work we will focus in particular on the dynamics of correlation and response functions, which are defined by 
\begin{align}
C(t,t') &= \frac12 \langle [\sigma(t),\sigma(t')]_{+} \rangle \\
R(t,t') &= \theta(t-t')\frac{i}{\hbar(t')}  \langle[\sigma(t),\sigma(t')]_{-} \rangle 
\end{align}
where $[A,B]_{\pm}=AB\pm BA$.  
The fully connected nature of the model defined in Eq.~(\ref{eqn:H}) allows us to derive closed dynamical equations that describe the evolution of correlation and response functions starting from an uncorrelated infinite temperature initial state.  After disorder-averaging and taking the $N \to \infty$ limit, the equations of motion for the correlation and response functions can be obtained following the method of Ref. \cite{Cugliandolo+98} and are given by 
{\medmuskip=-0.7mu
\thinmuskip=-0.7mu
\thickmuskip=-0.7mu
\begin{align}
& [ m \partial_t^2 + z(t)] R(t,t') = \delta(t-t') + \int_{0}^{\infty} \ud t'' \Sigma(t,t'') R(t'',t)
\label{eq.r} \\
&[ m \partial_t^2 + z(t)]  C(t,t') = \int_{0}^{\infty} \ud t'' \Sigma(t,t'')C(t'',t') \nn \\
& \quad \quad \quad\quad + \int_0^{t'}  \ud t'' D(t,t'') R(t',t'')
\label{eq.c}
\end{align}}
where we have defined the self-energies $\Sigma(t,t')$ and $D(t,t')$ as:
{\medmuskip=-0.5mu
\thinmuskip=-0.5mu
\thickmuskip=-0.5mu
\begin{align}
\Sigma(t,t') &= -\frac{p \mathcal{J}(t) \mathcal{J}(t')}{\hbar(t')} \textrm{Im} \left[ C(t,t') - \frac{i \hbar(t')}{2} R(t,t') \right]^{p-1} \label{eq.sig} \\
D(t,t') &= \frac{p \mathcal{J}(t) \mathcal{J}(t')}{2} \nn \\
 & \quad \times \textrm{Re} \left[ C(t,t')-\frac{i}{2}(\hbar(t') R(t,t') + \hbar(t) R(t',t)) \right]^{p-1}
\label{eq.D}
\end{align}}
With respect to the classical dynamical equations~\cite{Cugliandolo+17}, Eqs.~(\ref{eq.sig}-\ref{eq.D}) have extra self-energy contributions proportional to $\hbar(t)$ which arise from purely quantum fluctuations~\cite{,Cugliandolo+99}. We perform the dynamical evolution subject to a time-dependent Lagrange multiplier $z(t)$ used to enforce the global spherical constraint. We can derive the dynamical equation for this by taking the equal-time limit of Eq. \ref{eq.c} to obtain \cite{Cugliandolo+98}:
\begin{align}
z(t) &= \int_{0}^{t} \ud t'' \left[ \Sigma(t,t'')C(t'',t) + D(t,t'') R(t,t'') \right] \nn \\
& \quad \quad \quad - m \partial_t^2 \left. C(t,t') \right|_{t' \to t^{-}}
\label{eq.z}
\end{align}
Equations \ref{eq.r},\ref{eq.c} and \ref{eq.z} are the three dynamical equations whose solution will discuss in the remaining of the paper. Their causal structure allow for a simple discretisation and numerical solution- for further details, see the Supplementary Material \cite{SM}.

\emph{Finite Temperature Initial State Preparation and Double Quench -  }The dynamical equations~(\ref{eq.r},\ref{eq.c} and \ref{eq.z}) describe the evolution of the system from an initial infinite temperature initial state uncorrelated with the disorder. Here we are instead interested in studying dynamics from an initial finite temperature state, which would in principle require a three branch Keldysh contour structure as recently discussed~\cite{cugliandolo2019role}.  We instead perform the initial thermalisation numerically through a double-quench protocol. Specifically, we first quench from infinite temperature to some $T_0>T_d$ (where $T_d$ is the equilibrium dynamical temperature of the spin glass transition) and $\mathcal{J}(0<t<t_q) \equiv \mathcal{J}_0=1$ and $\hbar(0<t<t_q) \equiv \hbar_0$ and allow the system to thermalise in contact with a thermal bath, which we assume to be a set of harmonic oscillators in thermal equilibrium at some temperature $T_{0}$, as in Ref. \onlinecite{Cugliandolo+99}. This results in modifed self-energies $\tilde{\Sigma}(t,t')$ and $\tilde{D}(t,t')$ in Eq.~\ref{eq.D} due to the bath coupling, whose explicit expressions are given in~\cite{SM}. Then, for $t>t_q$ we switch off the coupling to the bath and let the system evolve unitarily with $\mathcal{J}(t \geq t_q) \equiv \mathcal{J}_F$ and $\hbar(t \geq t_q) \equiv \hbar_F$. All temperatures are measured in units of $\mathcal{J}_0$. Supporting data demonstrating that our system is well-equilibrated to the bath temperature is shown in Supplementary Material \cite{SM}. 

\begin{figure}[t!]
\begin{center}
\includegraphics[width= \linewidth]{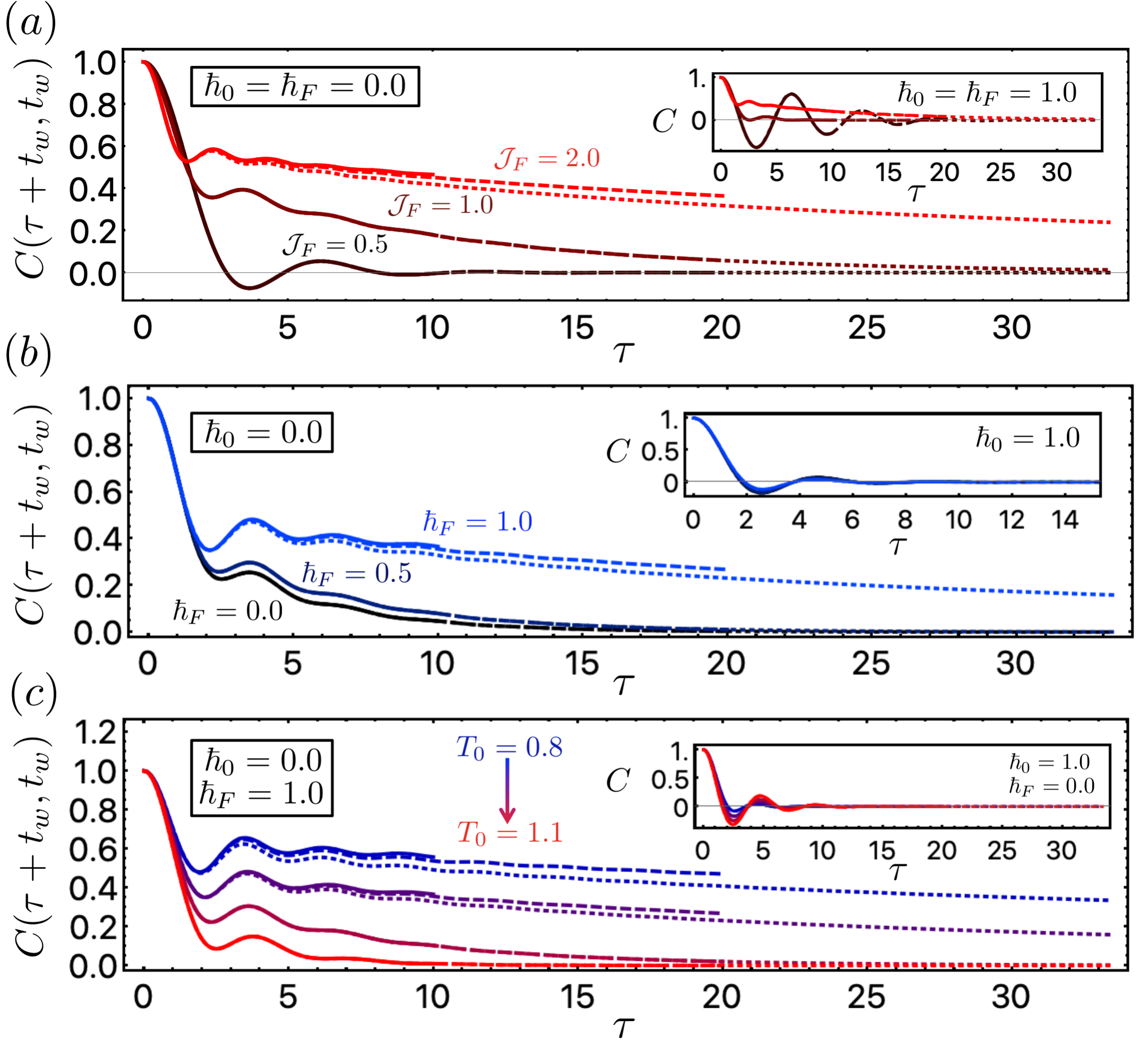}
\caption{Correlation functions after the second quench for a variety of different parameters, with $N=15000$ steps, $t_{max}=100$ and $t_q=t_{max}/2$. In each case, $\mathcal{J}_0=1.0$ and the wait times are $t_w=16.67$ (dotted), $t_w=30$ (dashed) and $t_w$=40 (solid).  a) Quench of  $\mathcal{J}$ in the classical (main panel) and quantum (inset) model, at  $T_{0}=0.8$.  Quenches with $\mathcal{J}_F<\mathcal{J}_0$ pump energy into the system, while quenches with $\mathcal{J}_F>\mathcal{J}_0$ extract energy and can lead to aging behaviour.
b) Quench of quantum fluctuations, for $T_{0}=0.9$ and $\mathcal{J}_F=1$.  A plateau emerges as $\hbar_F$ is increased starting from the classical limit. Vice versa, decreasing quantum fluctuations make the system thermalize rapidly (inset). 
c) Role of initial temperature, for $\mathcal{J}_F=\mathcal{J}_0=1$ and a sudden increase (main panel) or decrease (inset) of quantum fluctuations.  In the first case lowering the temperature leads to a dynamical glass transition, consistent with the shifted phase boundary of Figure 1, panel b)}
\label{fig.corr}
\end{center}
\end{figure}
\emph{Results -} For concreteness we will set $p=3$, though we expect our results to hold for any $p>2$.
 In Fig.~\ref{fig.corr} we plot the dynamics of correlation function $C(\tau+t_w,t_w)$ at fixed $\mathcal{J}_0=1$ for different type of quenches.  
We first study the dynamics keeping fixed the strength of quantum fluctuations while quenching $\mathcal{J}$ (panel a). We see that increasing $\mathcal{J}_F>\mathcal{J}_0$ results in a slow down of the dynamics and a plateau in the correlation function begins to emerge.  Such a plateau is associated with a non-zero Edwards-Anderson glassy order parameter. In the classical case $\hbar_F=\hbar_0=0$ we therefore recover the results of Ref.~\onlinecite{Cugliandolo+17}, while in the quantum case $\hbar_F=\hbar_0=1$  (see inset) we see that similar quenches of $\mathcal{J}$ does not lead to a well formed plateau, indicating that the quantum aging boundary shifts toward larger values of $\mathcal{J}_F/\mathcal{J}_0$. This is consistent with the naive expectation that quantum fluctuations suppress aging behavior. The resulting phase diagram is shown in Figure 1, panel a).
A rather different picture emerges instead when quantum fluctuations are suddenly quenched rather than kept fixed, as we show in panel b of Figure 2. Keeping the interaction fixed, $\mathcal{J}_F=\mathcal{J}_0$, and increasing the quantum fluctuations (main panel) strongly enhances the aging behavior of the system, as shown by the formation of a plateau and a waiting time dependence. On the contrary reducing the value of $\hbar_F<\hbar_0$ leads to a rapid relaxation (inset). This surprising outcome for a quench of $\hbar$ is further highlighted in panel c, where the dynamics for different initial temperature $T_0$ is studied. In particular we see that for an increase of $\hbar$ (main panel) the system upon cooling crosses a dynamical glass transition, even in absence of an interaction quench ($\mathcal{J}_F=\mathcal{J}_0$), and for temperature well above the classical $T_d$. On the contrary, decreasing $\hbar$ always keep the system in the paramagnetic phase. Those results therefore suggest that the aging regime is increased when quantum fluctuations are suddenly switched on, as we summarize in the panel b of Figure 1.

We remark that for the timescales accessible to our current simulations, the correlation function still decays and does not display a true plateau: this is likely an effect of not being able to access sufficiently long waiting times $t_w$ to see the true plateau, as evidenced by the strengthening of the plateau for larger $t_w$. By approximating $C(\tau+t_{w},t_{w})|_{\tau \to \infty}$ by the value of the correlation function at the longest times accessible to our simulation, and identifying this value with the Edwards-Anderson order parameter $q_{EA}$, we can plot an approximate non-equilibrium phase diagram for the isolated quantum system, shown in Fig. \ref{fig.phase}. Within our simulation times, as clearly shown by Fig. \ref{fig.corr}, we cannot reach the true $t \to \infty$ value of $q_{EA}$. Instead, we can set a threshold value and approximate that all $q \leq q_{th}$ are slowly decaying paramagnetic solutions, whereas for $q>q_{th}$ the system is in a true glassy phase. The results of this are shown in the phase diagram in Fig. \ref{fig.phase} by dashed lines, using $q_{th}\approx 0.2$, though the qualitative shape of the phase diagram does not depend strongly on this choice \footnote{To extract the true phase boundary one would need to consider the relaxation time of the system as a function of the control parameters and try to fit its divergence, however the timescales required to perform this analysis are longer than accessible with our numerical code.}.
\begin{figure}[t]
\begin{center}
\includegraphics[width= \linewidth]{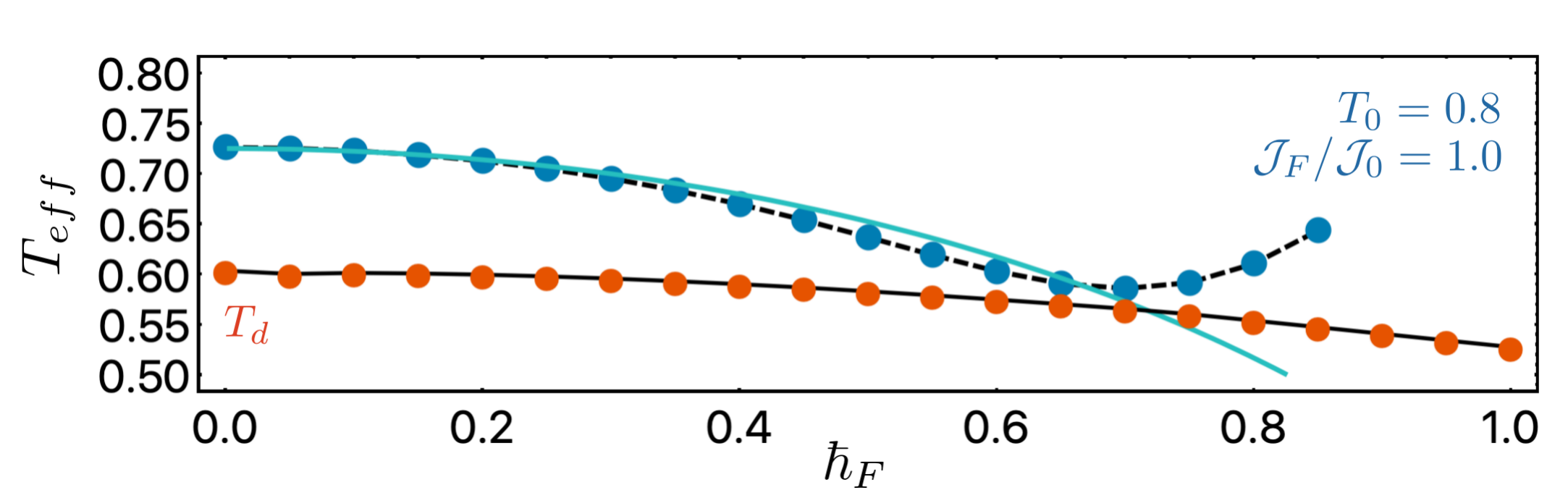}
\caption{
Effective temperature  $T_{eff}$ after the second quench for a system initially equilibrated at $T_{0}=0.8$, $\mathcal{J}_F = \mathcal{J}_0=1.0$ and $\hbar_0=0$, obtained from the dynamical equation and the fluctuation-dissipation relation (blue points). As the strength of quantum fluctuations is increased, $T_{eff}$ decreases until it reaches approximately the dynamical temperature $T_d$ (red points, obtained from dynamical simulations in the presence of a thermal bath), here at a value $\hbar_F \approx 0.7$. Beyond this, $T_{eff}$ displays the same non-monotonic behaviour seen in Ref. \onlinecite{Cugliandolo+17}, indicating a violation of FDT and suggesting that the system has entered the glass phase. For comparison, the \emph{thermodynamic} estimate for $T_{eff}$ (see main text) is shown as light blue line and matches almost perfectly the dynamical one. Other parameters $N=15000$ and $t_{max}=100$, with $t_q=t_{max}/2$. }
\label{fig.quenchT}
\end{center}
\end{figure}

\emph{Effective Temperature and Quench-Induced Cooling -} The results presented above indicate that quantum fluctuations and non-equilibrium effects can strongly enhance glassiness and increase the region of parameters where aging effects are observed. This is surprising at first, since glassiness is a low temperature property, while exciting the system with a global quantum quench injects extensive energy and should intuitively induce heating~\cite{MitraGiamarchiPRL11,SchiroMitraPRL14}. We can understand this effect in terms of an effective thermalisation of the isolated system to an effective temperature $T_{eff}$, as we show in detail by looking at the fluctuation-dissipation theorem (FDT) in the long-time regime of the dynamical equations for correlation and response~\cite{SM}. In Figure~\ref{fig.quenchT} we show that the $T_{eff}$ extracted from FDT \emph{decreases} with $\hbar_F$ and eventually reaches the dynamical critical temperature $T_d$ for the glass transition, below which the system fails to thermalise. By extracting the local minimum of $T_{eff}$ from Figure~\ref{fig.quenchT} and identifying it with the transition in our numerical data, we can draw a phase boundary with no free parameters, shown in Fig.  \ref{fig.phase} by the solid lines. Interestingly, the same effect of cooling by quantum fluctuations emerges from basic energetic arguments: indeed the effective temperature can be also estimated by comparing the post-quench energy $E_Q$, which is conserved during the unitary evolution, to the equilibrium internal energy of the system at a given value of $\hbar_F$, i.e. $E_Q=U(T_{eff},\hbar_F)$. Solving this equation for our model in the static approximation~\cite{Cugliandolo+01,SM}, which is valid in the high temperature phase under consideration, we obtain a \emph{thermodynamic} estimate for $T_{eff}$ which almost perfectly matches the dynamical one obtained from FDT in the regime where the system thermalises (see light blue line in Fig.~\ref{fig.quenchT}).

\emph{Discussion - } In our specific model~(\ref{eqn:H}) the strength of quantum fluctuations is controlled by the magnitude of $\hbar$. A natural question concerns whether the qualitative picture we presented so far would change in more realistic situations where quantum fluctuations are controlled by the action of a transverse field $\Gamma$, such as in the Ising p-spin quantum glass~\cite{GoldschmidtPRB90,Dobrosavljevic_1990,ObuchiEtAlJPS07,JorgEtAlPRL08}. In thermal equilibrium it is known that the spherical and the Ising p-spin share much of their physics~\cite{RitortIsingPspin,Cugliandolo+01,Cugliandolo+04,Takahashi_2011}, including the phase diagram which features a quantum glass to paramagnet phase transition driven by the strength of quantum fluctuations, encoded respectively in $\hbar$ or $\Gamma$. Whether this analogy remains valid also for the out of equilibrium dynamics is a priori not obvious. Using energetic arguments~\cite{SM} we estimate the effective temperature in the Ising p-spin after a quantum quench of the transverse field and show that, indeed, this quantity shows the same qualitative behavior in the two models. In particular we show that also in the Ising p-spin an increase of quantum fluctuation (i.e. a quench to a larger value of $\Gamma$) can lead to a decrease of the effective temperature, i.e. a cooling through quantum fluctuations that appears therefore a robust feature of isolated quantum glasses. This result is also of practical relevance, since quantum simulation of Ising $p$-spin models can be realized using arrays of superconducting qubits, which are modeled as two level systems with random Ising couplings and transverse fields, the latter tunable in real-time and therefore amenable to sudden or slow quenches. In fact, these protocols are routinely explored in the field of quantum annealing~\cite{BoixoEtalNatPhys2014}. Superconducting qubits also offer enough flexibility in fabrication and design such that arranging effective multi-spin interactions, such as those relevant for our p-spin with $p>2$ has indeed been already reported~\cite{PhysRevApplied.12.064026,melanson2019tunable}.

\emph{Conclusions - }
In this work we have studied the quench dynamics of an isolated quantum glass.
Remarkably, we have shown that suddenly increasing the strength of quantum fluctuations enhances aging behavior, in contradiction with common expectations based on the physics of quantum glasses coupled to thermal environment. The key feature of this effect relies on a `cooling by quantum fluctuations' effect that we have shown to hold also for the more realistic Ising p-spin case, a model which can be quantum simulated using superconducting qubits.

Interesting future directions include starting from a low temperature glass phase at $T<T_d$, for which the corresponding dynamical equations are already available in \cite{cugliandolo2019role}, to see how the quantum glasses respond to non equilibrium perturbations as well as to study the effect of a smooth quench protocol with finite duration, which may connect our results with investigations on quantum annealing done on related quantum glass models~\cite{JorgEtAlPRL08,J_rg_2010}. Solving the full real-time dynamics for other mean field models of isolated quantum glasses, such as the Ising $p$-spin and the quantum Random Energy Model, using similar techniques would also be an interesting direction to take.

We acknowledge helpful discussions with D. Abanin, G. Biroli,  L. Cugliandolo and M. Tarzia. This
work was supported by the grant “Investissements d'Avenir”
from LabEx PALM (ANR-10-LABX-0039-PALM), the grant DynDisQ from DIM SIRTEQ and by the
CNRS through the PICS-USA-14750. The majority of the computations were performed on the Coll\`ege de France IPH cluster computer.

\end{document}


\title{Supplementary Material to ``Quantum quenches in isolated quantum glasses out of equilibrium''}

\author{S. J. Thomson}
\email{steven.thomson@polytechnique.edu}
\affiliation{Centre de Physique Th\'{e}orique, CNRS, Institut Polytechnique de Paris, Route de Saclay, F-91128 Palaiseau, France}
\affiliation{Institut de Physique Th\'{e}orique, Universit\'{e} Paris Saclay, CNRS, CEA, F-91191 Gif-sur-Yvette, France}
\author{P. Urbani}
\email{pierfrancesco.urbani@ipht.fr}
\affiliation{Institut de Physique Th\'{e}orique, Universit\'{e} Paris Saclay, CNRS, CEA, F-91191 Gif-sur-Yvette, France}
\author{M. Schir\'o}
\email{marco.schiro@ipht.fr}
\thanks{ On Leave from: Institut de Physique Th\'{e}orique, Universit\'{e} Paris Saclay, CNRS, CEA, F-91191 Gif-sur-Yvette, France}
\affiliation{JEIP, USR 3573 CNRS, Coll\`ege de France,   PSL  University, 11,  place  Marcelin  Berthelot,75231 Paris Cedex 05, France}
\date{\today}

\begin{abstract}
In this Supplementary Material, we provide additional technical details about the derivation of the dynamical equations, the numerical procedure used to solve the dynamical equations and additional data confirming that the initial equilibration phase of our double-quench procedure performs as expected.
\end{abstract}

\maketitle

\section{Derivation of the Dynamical Equations}

\subsection{Linear Response}
In order to allow the strength of quantum fluctuations in this calculation to be parametrised by a time-dependent $\hbar(t)$, we must be precise about how this time-dependence enters in the various quantities which we are interested in computing. In particular, to define the response function in the presence of a time-dependent $\hbar(t)$, we must first revisit linear response theory. Defining the Schr\"odinger equation as $i \hbar(t) \partial_t \psi(t) = \hat{\mathcal{H}} \psi(t)$, linear response theory leads to the definition:
\begin{align}
R(t,t') = \frac{i}{\hbar(t')} \Theta(t-t') \langle [O(t),O(t')] \rangle 
\end{align}
where the $\hbar(t')$ in the response function takes the earlier of the two time arguments, and $O(t)$ represents an arbitrary local observable. This will be important in evaluating the saddle-point equations of the disorder-averaged action.

\subsection{Generating Functional}
The generating functional can be written as:
\begin{align}
\mathcal{Z}[\xi^+,\xi^-] &= \int \mathcal{D}\sigma^+ \mathcal{D} \sigma^- \exp \left[ i \left( \tilde{S}[\sigma^+] - \tilde{S}[\sigma^-]  + \int \frac{\ud t}{\hbar(t)} (\xi^+(t)\sigma^+(t) - \xi^-(t) \sigma^-(t)) \right) \right] \braket{\sigma+| \rho(0) | \sigma^-}
\end{align}
where $\rho(0)$ represents the element of the initial density matrix at $t=0$ and is chosen to be uncorrelated with the disorder (i.e. a random, infinite-temperature initial state). The tilde notation signifies that the time-dependent factors of $1/\hbar(t)$ have been absorbed into the definition of the action and appear under the relevant time integrals. The action $\tilde{S}$ can be defined in terms of a quadratic term $S_0$ and a disordered interaction term $V[\bm{\sigma},J]$:
\begin{align}
\tilde{S}_{s} [\bm{\sigma}, J] &= \tilde{S}_0[\bm{\sigma}]  - \int_0^{\infty} \frac{\ud t}{\hbar(t)} \phantom{.} V[\bm{\sigma}, J], \\
\tilde{S}_0[\bm{\sigma}]  &= \int_0^{\infty} \frac{\ud t}{\hbar(t)} \left[ \frac{m}{2} \dot{\bm{\sigma}}^{2} - \frac{z}{2} (\bm{\sigma}^2 - N)\right], \\
V[\bm{\sigma}, J] &= \mathcal{J}(t)\sum_{i_1 < ... < i_p}^{N} J_{i_1...i_p} \sigma_{i_1}...\sigma_{i_p}.
\end{align}
The quadratic part $S_0$ contains a kinetic term, chosen such that the eventual dynamical equations are written in terms of second derivatives with respect to time. In the presence of the global spherical constraint that enforces $\sum_{i}^{N} \sigma_i^2 = N$ through the introduction of a Lagrange multiplier $z$, the variables $\sigma_i$ are real, continous dynamic variables constrained to lie between $-\sqrt{N} < \sigma_i < \sqrt{N} \phantom{.} \forall i$.

This action can be split into components $\bm{\sigma}^{+}$ and $\bm{\sigma}^{-}$ residing on the forward and backwards Keldysh contours respectively, to give an action:
\begin{align}
\tilde{S}[\bm{\sigma}^{+},\bm{\sigma}^{-}, J]  &= \tilde{S}_0[\bm{\sigma}^{+}] - \tilde{S}_0[\bm{\sigma}^{-}] - \int_0^{\infty} \frac{\ud t}{\hbar(t)} \left( V[\bm{\sigma}^{+}, J]  - V[\bm{\sigma}^{-}, J]  \right)
\end{align}
where the relative minus sign comes from reversing the integration limits on the reverse contour. For clarity, in the following we will drop the tilde notation, but it is to be understood that the time-dependent $\hbar(t)$ remains under the relevant integrals in all of the following expressions.

\subsection{System-Bath Coupling}

The coupling between the system and bath can be treated exactly as in Ref. \onlinecite{Cugliandolo+99}. The time-dependent $\hbar(t)$ presents no complications, as we switch off the system-bath coupling at a time $t_q$, defined as a time where the system has initially equilibrated with the thermal bath. Therefore, all factors of $\hbar(t)$ can simply be replaced by $\hbar_0$ in the following expressions. We couple the quantum $p$-spin Hamiltonian linearly to a bath of harmonic oscilllators assumed to be in thermal equilibrium. This coupling can be described by the Feynman-Vernon influence functional:
\begin{align}
S_{bath} = \frac{1}{\hbar_0} \int_0^{t_q} \ud t \int_0^{t_q} \ud t' \left( -[\bm{\sigma}^+(t)-\bm{\sigma}^-(t)] \eta(t-t') [\bm{\sigma}^+(t') + \bm{\sigma}^-(t')] + i [\bm{\sigma}^+(t) - \bm{\sigma}^-(t)] \nu(t-t')[\bm{\sigma}^+(t') - \bm{\sigma}^-(t')] \right)
\end{align}
where $\eta$ and $\nu$ are the correlation and response functions of the bath, and are time-translation invariant due to the bath being in equilibrium. They are given by:
\begin{align}
\eta(t-t') &= -\Theta(t-t') \int_0^{\infty} \ud \omega I(\omega) \sin [\omega(t-t')], \\
\nu(t-t') &= \int_0^{\infty} I(\omega) \coth \left( \frac12 \beta \hbar_0 \omega \right) \cos[\omega(t-t')]
\end{align}
where $I(\omega)$ is the spectral function of the bath. We choose an Ohmic bath with $I(\omega) = \frac{1}{\pi} \exp \left( -|\omega|/\Lambda \right)$, for which explicit expressions for both $\nu(t-t')$ and $\eta(t-t')$ can be straightfowardly found, and we choose the integration cutoff to be $\Lambda=5$. 

\subsection{Disorder Averaging}
Under the assumption that $\hat{\rho}(0)$ is uncorrelated with the disorder, we can perform the disorder average explicitly. Assuming the distribution of the disorder variable $J_{i_1...i_p}$ is given by a Gaussian with zero mean and variance of the form:
\begin{align}
P[J] &= \sqrt{\frac{N^{p-1}}{ \pi p!}} \exp \left( - \frac{N^{p-1}}{p!} \sum_{i_1 \neq ... \neq i_p} (J_{i_1...i_p})^{2} \right) \label{eq.pdist}
\end{align}
such that
\begin{align}
\overline{(J_{i_1...i_p})^{2} } =  \int \ud J P[J] (J_{i_1...i_p})^{2}  = \frac{p!}{2 N^{p-1}}
\label{eq.avgJ}
\end{align}
where the factors of $N$ and $p$ are included to simplify later expressions. The prefactor $\mathcal{J}(t)$ shown in the main text therefore acts as a a time-dependent disorder strength that we can use in the double-quench procedure. The average of any operator is given by:
\begin{align}
\overline{\langle \bm{\sigma}(t) \rangle} & \equiv \left. \overline{\frac{\partial \ln \mathcal{Z}[\bm{\xi}^{+},\bm{\xi}^{-},J]}{\partial \bm{\xi}^{+}(t)}} \right|_{\bm{\xi}=0} = \frac{1}{\mathcal{Z}[0,0,J]}  \left. \frac{\partial \overline{ \mathcal{Z}[\bm{\xi}^{+},\bm{\xi}^{-},J]}}{\partial \bm{\xi}^{+}(t)} \right|_{\bm{\xi}=0}
\end{align}
from which it follows that the only averaging to be done is over $\mathcal{Z}$ itself rather than $\ln \mathcal{Z}$. The disorder average therefore reduces to just averaging over the $p$-spin vertex, the only term to have a disorder dependence. It can be performed explicitly to give a disorder-averaged generating functional:
\begin{align} 
&\overline{ \mathcal{Z}[\bm{\xi}^{+},\bm{\xi}^{-},J]} = \int \mathcal{D}_{\bm{\sigma}^{-}} \mathcal{D}_{\bm{\sigma}^{+}} \exp\left[ i \left( S_{eff}[\bm{\sigma^{+}},\bm{\sigma^{}}] + \int \ud t (\bm{\xi}^{+}(t)\bm{\sigma}^{+}(t)-\bm{\xi}^{-}(t)\bm{\sigma}^{-}(t))  \right) \right]
\end{align}
with the disorder-averaged effective action:
\begin{align}
&S_{eff}[\bm{\sigma^{+}},\bm{\sigma^{}}]  = S_0[\bm{\sigma}^{+}] - S_0[\bm{\sigma}^{-}] - V_{D}[\bm{\sigma}^{+},\bm{\sigma}^{-}] +S_{bath}[\bm{\sigma^{+}},\bm{\sigma^{}}], \\
&V_{D}[\bm{\sigma}^{+},\bm{\sigma}^{-}] = \frac{i N}{4}  \int \ud t \ud t' \frac{ \mathcal{J}(t) \mathcal{J}(t')}{ \hbar(t) \hbar(t')} \sum_{\alpha,\beta = \pm} \alpha \beta \left( \frac{1}{N}\bm{\sigma}^{\alpha} (t)\bm{\sigma}^{\beta}(t') \right)^{p} 
\end{align}
where $\alpha,\beta=\pm$ are the Schwinger-Keldysh contours.

\subsection{Transformed Order Parameters}

The quadratic part of the action may be written in matrix form as:
\begin{align}
S_{eff}^{(2)}[\bm{\sigma}^{+},\bm{\sigma}^{-}] &= -\frac12 \int \ud t  \ud t'\bm{\sigma}^{\alpha}(t) O_{p}^{\alpha \beta}(t,t') \bm{\sigma}^{\beta}(t')
\end{align}
where the matrix elements are given by:
\begin{align}
O_{p}^{++}(t,t') &= \frac{1}{\hbar(t)}[m \partial_t^{2} + z^{+}(t)] \delta(t-t') - \frac{2}{\hbar_0} (i\nu(t-t')-\eta(t-t')) \Theta(t-t_q) \Theta(t'-t_q), \\
O_{p}^{+-}(t,t') &= \frac{1}{\hbar_0} (2 \eta(t-t') + 2 i \nu(t-t'))\Theta(t-t_q) \Theta(t'-t_q), \\
O_{p}^{-+}(t,t') &= \frac{1}{\hbar_0} (-2 \eta(t-t') + 2 i \nu(t-t'))\Theta(t-t_q) \Theta(t'-t_q), \\
O_{p}^{--}(t,t') &= -\frac{1}{\hbar(t)}[m \partial_t^{2} + z^{-}(t)] \delta(t-t')- \frac{2}{\hbar_0} (i\nu(t-t')+ \eta(t-t'))\Theta(t-t_q) \Theta(t'-t_q).
\end{align}
We now introduce new variables $Q^{\alpha \beta}(t,t')$ (where $\alpha,\beta=\pm$) which will allow us to decouple the $p$-interaction term:
\begin{align}
1 &= \int \prod_{\alpha \beta} \mathcal{D} Q^{\alpha \beta} \delta \left( \frac{1}{N} \bm{\sigma}^{\alpha}(t) \bm{\sigma}^{\beta}(t') - Q^{\alpha \beta}(t,t')\right), \\ 
& \propto \int \prod_{\alpha \beta} \mathcal{D} Q^{\alpha \beta} \mathcal{D} \lambda^{\alpha \beta} \exp \left( -\frac{i}{2} \lambda^{\alpha \beta} \left( \bm{\sigma}^{\alpha}(t) \bm{\sigma}^{\beta}(t') - N Q^{\alpha \beta}(t,t')\right) \right).
\end{align} 

\subsection{Saddle-Point Equations}
Following the notation of Ref. \onlinecite{Cugliandolo+99}, we can define the following matrices:
\begin{align}
\Lambda(t,t') = \begin{bmatrix} \lambda^{++}(t,t') & \lambda^{+-}(t,t') \\ \lambda{-+}(t,t') & \lambda^{--}(t,t') \end{bmatrix}, \quad
\mathcal{Q}(t,t') = \begin{bmatrix} Q^{++}(t,t') & Q^{+-} (t,t')\\ Q{-+}(t,t') & Q^{--}(t,t') \end{bmatrix} 
\end{align}
to allow us to compactly encode all correlations between contours. Further, since taking derivatives of the action with respect to $Q$ will naturally lead to terms of the form $Q^{p-1}$, we also need to define the matrix:
\begin{align}
F[Q](t,t') &= \begin{bmatrix} [Q^{++} (t,t')]^{p-1} & -[Q^{+-} (t,t')]^{p-1} \\ -[Q^{-+} (t,t')]^{p-1} & [Q^{--} (t,t')]^{p-1} \end{bmatrix}  \frac{\mathcal{J}(t) \mathcal{J}(t')}{\hbar(t) \hbar(t')}
\end{align}
where the time-dependent parameters $\mathcal{J}(t)$ and $\hbar(t)$ are included in the definition of $F[Q]$ as it is necessary to keep track of these time-dependences throughout the following procedure. In this notation, the saddle point equation with respect to $\Lambda(t,t')$ is given by:
\begin{align}
i \Lambda(t,t') = \mathcal{Q}^{-1}(t,t') - iO_p(t,t')
\end{align}
And with respect to $\mathcal{Q}(t,t')$:
\begin{align}
i\Lambda(t,t') = \frac{p}{2} F[Q](t,t')
\end{align}
We can combine these two saddle-point equations to eliminate the variable $\Lambda$ and obtain:
\begin{align}
\mathcal{Q}^{-1}(t,t') - i O_p(t,t') = \frac{ p}{2} F[Q](t,t')
\end{align}
Multiplying from the right with the matrix $\mathcal{Q}$ results in the final equation:
\begin{align}
i O_p \otimes \mathcal{Q} (t,t')=  \mathcal{I} - \frac{p}{2} F[Q] \otimes \mathcal{Q}(t,t') \label{eq.sp}
\end{align}
where the $\otimes$ symbol denotes the operational product and we have defined $\mathcal{I} = \mathcal{I}^{\alpha \beta}(t,t') = \delta_{\alpha \beta} \delta(t-t')$. The saddle-point equations with respect to the Lagrange multiplier $z^{\alpha}$ reduce to the definition of the spherical constraint on both contours, and are not shown explicitly here.

\subsection{Dynamical Equations}
 
We can express both $C(t,t')$ and $R(t,t')$ in terms of appropriate sums over $Q^{\alpha \beta}(t,t')$, and so the dynamical equations for these variables can be obtained from the saddle-point equations. Specifically:
\begin{align}
R(t,t') = \frac{i}{\hbar(t')} [Q^{++}(t,t') - Q^{+-}(t,t')]
\end{align}
and so the dynamical equation for the response function follows from subtracting the $++$ and $+-$ components of Eq. \ref{eq.sp}. Note that the time-dependent $\hbar(t)$ presents a complication with respect to the derivation of Ref. \onlinecite{Cugliandolo+99}, and that adding together these components of Eq. \ref{eq.sp} gives:
\begin{align}
\left( \frac{\hbar(t')}{\hbar(t)} \right) [ m \partial_t^2 + z(t)] R(t,t') &= \delta(t-t') +  \left( \frac{\hbar(t')}{\hbar(t)} \right) \int_{0}^{t} \ud t'' \Sigma(t,t'') R(t'',t')
\end{align}
where the self-energy $\Sigma(t,t')$ is defined below. Multiplying through by $\hbar(t)/\hbar(t')$ and using that $(\hbar(t)/\hbar(t')) \delta(t-t') = \delta(t-t')$, we recover the same dynamical equation as in Ref. \onlinecite{Cugliandolo+99}. Similarly, the correlation function is given by:
\begin{align}
C(t,t') = \frac12 [Q^{+-}(t,t') + Q^{-+}(t,t')]
\end{align}
the dynamical equation for which is obtained from the addition of the $+-$ and $-+$ components of Eq. \ref{eq.sp}. The detailed derivation follows the steps of Ref. \onlinecite{Cugliandolo+99} with no additional complications and as such is not reproduced here.

\subsection{Final Dynamical Equations}

We can write both dynamical equations in a compact form as:
\begin{align}
[ m \partial_t^2 + z(t)] R(t,t') &= \delta(t-t') + \int_{0}^{t} \ud t'' \Sigma(t,t'') R(t'',t') \label{eq.r}, \\
[ m \partial_t^2 + z(t)]  C(t,t') &= \int_{0}^{t} \ud t'' \Sigma(t,t'')C(t'',t') + \int_0^{t'}  \ud t'' D(t,t'') R(t',t'') \label{eq.corr}
\end{align}
where we have defined the self-energy $\Sigma(t,t')$ and the vertex $D(t,t')$ as the following:
\begin{align}
\Sigma(t,t') &= -4 \eta(t-t') \Theta(t_q-t)\Theta(t_q-t')  -\frac{p \mathcal{J}(t) \mathcal{J}(t')}{\hbar(t')} \textrm{Im} \left[ C(t,t') - \frac{i \hbar(t')}{2} R(t,t') \right]^{p-1}, \\
D(t,t') &= 2 \hbar_0 \nu(t-t') \Theta(t_q-t)\Theta(t_q-t') + \frac{p \mathcal{J}(t) \mathcal{J}(t')}{2} \textrm{Re} \left[ C(t,t')-\frac{i}{2}(\hbar(t') R(t,t') + \hbar(t) R(t',t)) \right]^{p-1}.
\end{align}
From the structure of the saddle-point equations, we can also obtain the equal-time relations:
\begin{align}
C(t,t) &= 1 \quad \quad \quad \quad \partial_t C(t,t') |_{t' \to t^-}  = \partial_t C(t,t') |_{t' \to t^+}  = 0 \\
R(t,t) &= 0 \quad \quad \quad \quad \partial_t R(t,t') |_{t' \to t^-} = \frac{1}{m} \quad \quad \quad \quad \partial_t R(t,t') |_{t' \to t^+} = 0.
\end{align}
At this point, we have the equations for the evolution of the correlation and response functions, but we need to perform this evolution subject to the time-dependent Lagrange multiplier $z(t)$ which is used to enforce the global spherical constraint at all times. We obtain the dynamical equation for $z(t)$ by taking the $t' \to t^-$ limit of Eq. \ref{eq.corr} above:
\begin{align}
z(t) &= \int_{0}^{t} \ud t'' \left[ \Sigma(t,t'')C(t'',t) + D(t,t'') R(t,t'') \right] - m \partial_t^2 \left. C(t,t') \right|_{t' \to t^{-}}
\end{align}

\section{Numerical Considerations}

The equations of motion can be straightforwardly discretised and solved numerically, however for an isolated system numerical errors can lead to a violation of energy conservation, with an error that increases in time proportionally with the step size $\delta t$. This error can prevent the numerical algorithm from reaching long times after the second quench. We overcome this problem by following the procedure of Ref. \onlinecite{Cugliandolo+17} and replacing the second-derivative term in $z(t)$ by an energy density:
\begin{align}
&z(t) = \frac{(p+2)}{p} \int_{0}^{t} \ud t'' \left[ \Sigma(t,t'')C(t'',t) + D(t,t'') R(t,t'') \right] +2E(t)
\end{align}
where $E(t) = E_{k}(t) + E_p(t)$ is the total energy density of the system. For the isolated system, we can fix $E(t)$ immediately following the second quench, which enforces energy conservation and allows us to reach long times while maintaining reasonable numerical accuracy. This can be derived in the following way. The kinetic energy is given by:
\begin{align}
E_k &= \frac{m}{2} \sum_i \langle \dot{\sigma}_i(t)^{2} \rangle = -\frac{m}{2}  \left. \partial_{t}^{2} C(t,t') \right|_{t' \rightarrow t^{-}}
\end{align}
as discussed in Ref \onlinecite{Cugliandolo+17}. The potential energy density\footnote{Note that in the case of a system-bath coupling, the bath contributions must be subtracted from the self-energy $\Sigma(t,t')$ and vertex $D(t,t')$ shown here.} is given by\cite{Cugliandolo+04b}:
\begin{align}
E_p &= -\frac{1}{N} \left[ \left\langle \mathcal{J}(t) \sum_{i_1...i_p} J_{i_1...i_p} \sigma_{i_1}(t)...\sigma_{i_p}(t) \right\rangle \right] =\frac{1}{p} \int_0^{\infty} \ud t' \left[ \Sigma(t,t') C(t,t') + D(t,t') R(t,t')  \right]
\end{align}
and by combining these two expressions and using that the total energy is $E(t) = E_k(t) + E_p(t)$, we can rewrite the Lagrange multiplier and obtain the expression above. For the classical system, we checked numerically that these quantities reproduce known classical results. For the quantum system, we verified that the resulting energy matches that obtained with free unconstrained time-evolution for short periods following the second quench, before the numerical error becomes significant.

\section{Fluctuation-Dissipation Theorem}

\begin{figure}[t!]
\begin{center}
\includegraphics[width= 0.65\linewidth]{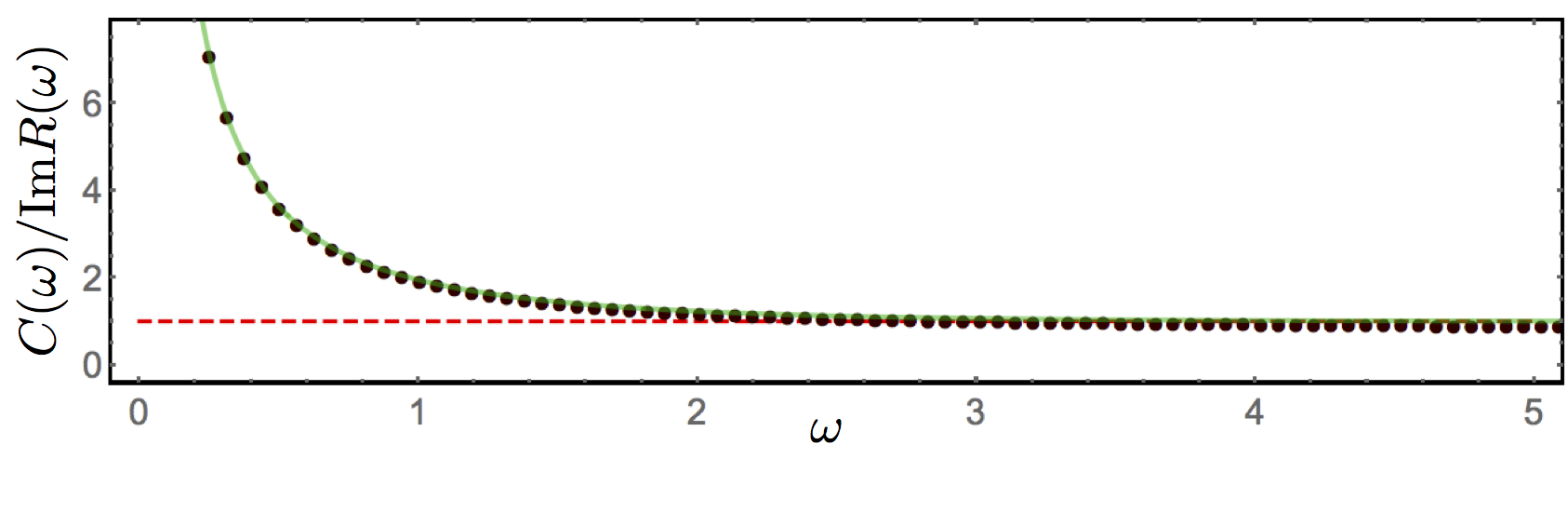}
\caption{The ratio $C(t,\omega)/\textrm{Im} \left[ R(t,\omega) \right]$ computed numerically (black dots) and fitted with a function of the form shown in Eq. \ref{eq.fdtfreq} (green solid line) for a system coupled to a bath with $t_{max}=100$, $dt = 0.006\dot{6}$, $\mathcal{J}=1$,  $T_{bath} = 1.0$ and $\hbar=1.0$. The red dashed line is a guide to the eye at $\lim_{y \to \infty}\coth{y}=1$. We see that the numerical data is very well fitted by the $\coth$ function and quantum FDT holds.}
\label{fig.fdtfreq}
\end{center}
\end{figure}

The quantum fluctuation-dissipation relation for an equilibrium system is:
\begin{align}
R(\tau) &= \frac{2i}{\hbar} \Theta(\tau) \int_{-\infty}^{\infty} \frac{\ud \omega}{2 \pi} \exp[-i \omega \tau] \tanh \left( \frac{\beta \hbar \omega}{2} \right) C(\omega)
\end{align}
where $C(\omega)$ is the Fourier transform of the (time-translation invariant) correlation function $C(\tau)$ with $\tau=t-t'$. In the case where $\beta \hbar \omega/2 \ll 1$, i.e. in the limits of high temperature, low frequency or $\hbar \to 0$, we recover the classical result:
\begin{align}
R(\tau) &= -\frac{1}{T} \frac{\ud C(\tau)}{\ud \tau}.
\label{eq.fdt1}
\end{align}
The generalised quantum fluctuation-dissipation theorem for a system out of equilibrium\cite{Cugliandolo+99} is given by:
\begin{align}
R(t,t') &= \frac{2i}{\hbar} \Theta(t-t') \int_{-\infty}^{\infty} \frac{\ud \omega}{2 \pi} \exp[-i \omega (t-t')] \tanh \left( \frac{\beta_{eff} \hbar \omega}{2} \right) C(t,\omega)
\end{align}
where $\beta_{eff}=1/T_{eff}$ is the inverse effective temperature and
\begin{align}
C(t,\omega) = 2 \textrm{Re} \int_0^t \ud t'' \exp[i \omega (t-t'')] C(t,t'')
\end{align}
is a partial Fourier transform of the correlation function. In principle, the effective temperature is a function of both times $t$ and $t'$ as well as $\hbar(t)$, however we suppress this dependence for clarity of presentation. 

To check that our system is well-equilibrated before we cut off the bath and perform the second quench, we can compute the Fourier-transformed correlation and response. From the above quantum FDT relation, the ratio of these is given by:
\begin{align}
\frac{C(t,\omega)}{\textrm{Im} \left[ R(t,\omega) \right]} = \hbar \coth\left( \frac{\beta^* \hbar \omega}{2} \right)
\label{eq.fdtfreq}
\end{align}
We can compute this ratio numerically and fit it with a $\coth$ function to extract the temperature of the system: results are shown in Fig. \ref{fig.fdtfreq}. We see very good agreement between the quantum FDT and the numerical data, confirming that in equilibrium, our numerical procedure behaves as expected.

After the second quench, the system is manifestly out of equilibrium and the FDT only holds for the low-frequency degrees of freedom. Following the results of Ref. \onlinecite{Cugliandolo+99} who showed that the correlation and response of the quantum system in the aging regime are related in a remarkably similar way to the classical FDT, we use a generalisation of the classical result (Eq. \ref{eq.fdt1}) with $T \to T_{eff}$. To extract this temperature, it is convenient to define the integrated response:
\begin{align}
\chi(t,t') &= \int_{t'}^t \ud t'' R(t, t'').
\end{align}
For a classical system in equilibrium, a plot of $\chi(t,t')$ against $C(t,t')$ will be a straight line with gradient $-1/T$. Violations of FDT will result in a deviation away from a straight line and often a visible separation of time scales, however we can still use this relation to extract an effective temperature in the long-time limit.

\pagebreak
\section{Equilibration}

\subsection{Equilibration with a bath}

\begin{figure}[t!]
\begin{center}
\includegraphics[width= 0.8\linewidth]{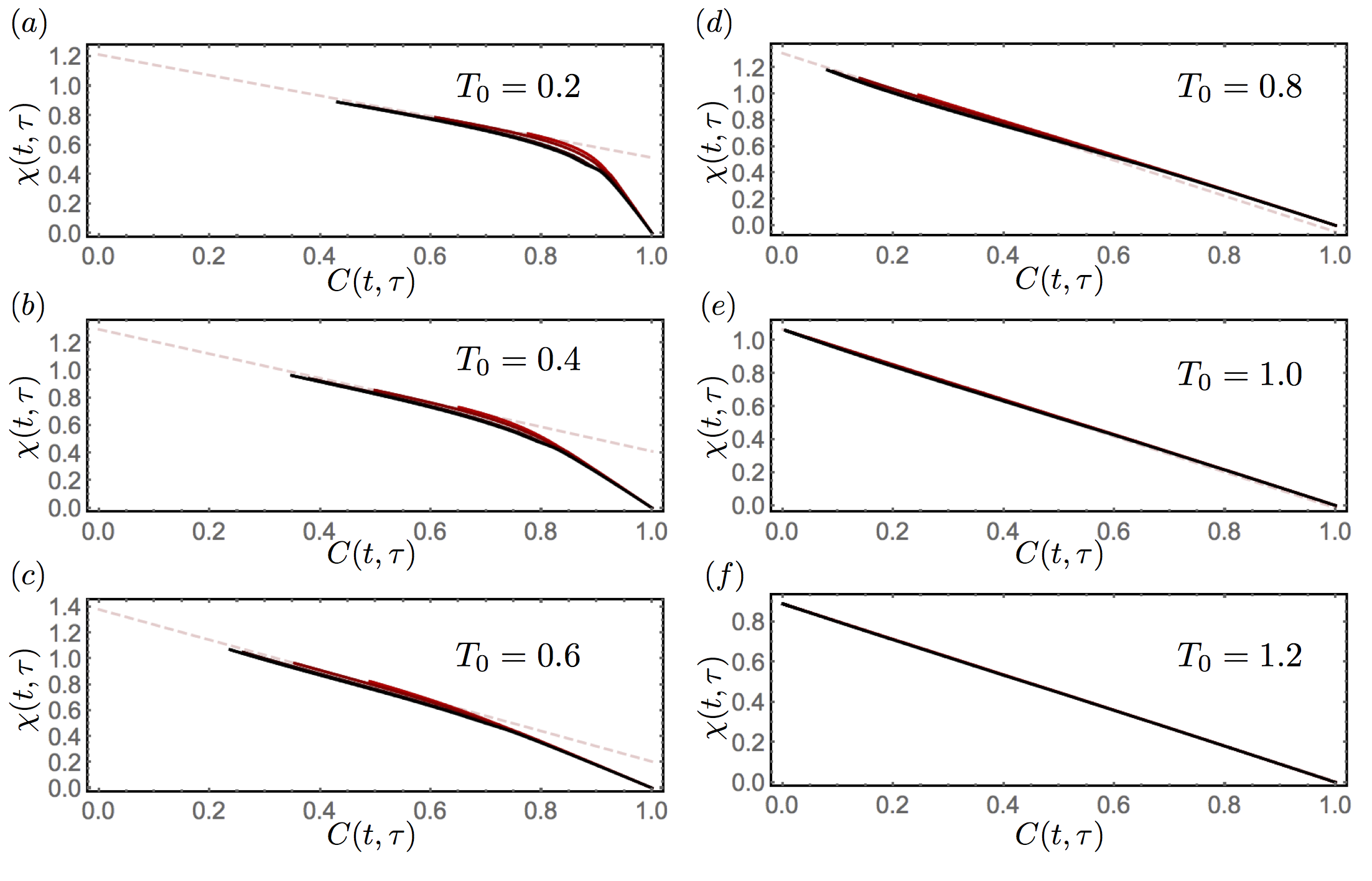}
\caption{The integrated response $\chi(t,\tau)$ of the classical $p$-spin model plotted parametrically against the correlation function $C(t,\tau)$ for a variety of $\tau = 5, 6.25. 12.5, 20$ (from black to red resepectively). In the left column, the system is in the glass phase: time-translation invariance is broken and the FDT shows the clear existence of two time scales, each characterised by a different effective temperature (i.e. slope of the line). In the right column, the system is in the paramagnetic phase: the plots display time-translation invariance and the curves lie on top of one another. At $T_{0}=0.8$, the system is paramagnetic but close to the transition: the dynamics in this regime are extremely slow, reflected by the waiting-time dependence starting to emerge at long times. All data were taken with $t_{max}=50$ and $\delta t = 0.006\dot{6}$.}
\label{fig.fdt}
\end{center}
\end{figure}

In order to establish the accuracy of the double-quench protocol, we first test the quality of the initial equilibration step by using the FDT to extract the temperature of the system in the time window $0<t<t_q$ and check that it reproduces known features of the equilibrium system. The interated response plotted against the correlation function for a variety of wait times and bath temperatures are shown in Fig. \ref{fig.fdt}, in the well-studied classical case of $\hbar_0 = 0$. We see that in the paramagnetic phase, the system displays time-translation invariance (TTI) and obeys classical equilibrium FDT, whereas at low temperature, the FDT is violated and there is a clear separation of time scales: an early-time transient regime which displays time-translational invariance, followed by the onset of an aging regime where TTI is broken and the system is no longer in equilibrium.

\subsection{Effective Temperature}

\begin{figure}[t!]
\begin{center}
\includegraphics[width= 0.65\linewidth]{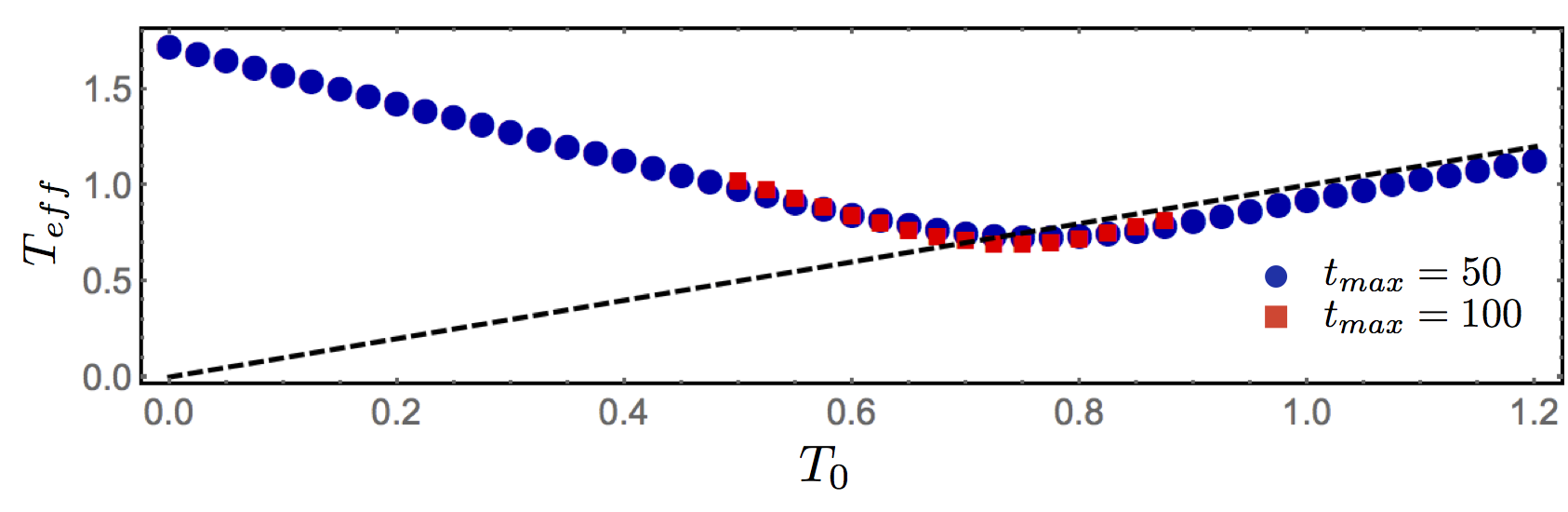}
\caption{The effective temperature $T_{eff}$ obtained from the FDT relation plotted against the bath temperature $T_{0}$. The black dashed line is a guide to the eye with slope equal to one. Both sets of data points use the same step size ($\delta t = 0.006\dot{6}$). Note the non-monotonic behaviour of $T_{eff}$ on crossing the phase transition from above: this feature can be used to identify a transition temperature from the numerical data.}
\label{fig.Teff}
\end{center}
\end{figure}

In Fig. \ref{fig.Teff}, we show the effective temperature extracted from the equilibrium FDT relation plotted against the bath temperature, for two different maximum times $t_{max}=50$ (with $N=7500$) and $t_{max}=100$ ($N=15,000$) in blue and red respectively. We see that the dependence on $t_{max}$ is slight, giving us good confidence that the pre-quench equilibration used in the main text is sufficiently accurate. In the paramagnetic phase, the error $\Delta T = T_{0} - T_{eff}$ is controlled predominantly by the step-size $\delta t$ used - we typically use $\delta t = 0.006\dot{6}$ in order to obtain a good compromise between retaining numerical accuracy and being able to reach a suitably large $t_{max}$. 
\pagebreak

A key feature of Fig. \ref{fig.Teff} is the non-monotonic behaviour of the effective temperature $T_{eff}$, which rises sharply as the bath temperature decreases and we pass through the phase transition. This behaviour has been seen in a variety of other works, most notably Ref. \onlinecite{Cugliandolo+17}. In the present work, we take the minimum value of the effective temperature $T^{*}_{eff}$ to indicate the phase transition in our numerical data; this is how we compute the equilibrium phase boundary shown in the main text. The true temperature of the equilibrium phase transition in the classical $p$-spin model is $T_{d} \approx 0.6$. Due to the finite timescale of our simulation and the increasingly slow dynamics on approach to the transition, the transition in our data occurs at a slightly higher bath temperature: this is symptomatic of the difficulties in extracting a glassy phase transition from dynamical data taken on finite timescales. We expect this finite-time error to result in a small systematic shift of all of our results, but to have no qualitative effect.

\subsection{Equilibration of the isolated system}

\begin{figure}[h!]
\begin{center}
\includegraphics[width= 0.8\linewidth]{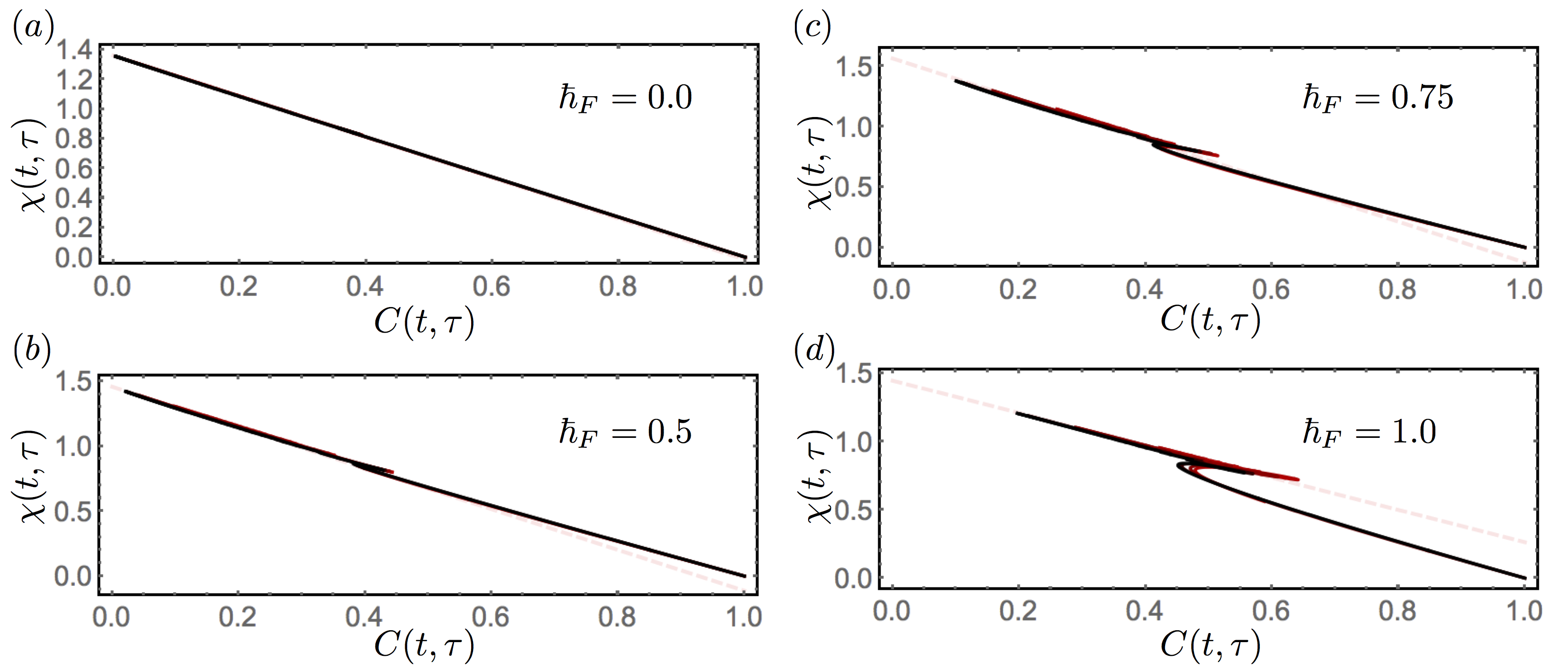}
\caption{The same quantities as Fig. \ref{fig.fdt}, but now after the second quench. In each case, the system was first equilibrated at a temperature $T_{0}=0.8$ with $\hbar_0=0.0$. For small $\hbar_F$, the system continues to respect the classical FDT and the plots are approximately straight lines, however as $\hbar_F$ increases above the transition at approximately $\hbar_F \approx 0.7$, we see that the system violates classical FDT and the parametric plots take a highly non-trivial form with two distinct time sectors.}
\label{fig.fdt2}
\end{center}
\end{figure}

After the second quench, the system is out of equilibrium but we can still use the generalised FDT in the low-frequency limit to extract an effective temperature. Representative plots of the integrated response for a system equilibrated with a classical bath at $T_{0}=0.8$ are shown in Fig. \ref{fig.fdt2}. For small quenches in $\hbar_F$, the system remains in equilibrium and continues to obey the classical FDT, as evidenced by Fig. \ref{fig.fdt2}(a). As we increase $\hbar_F$, the system enters an aging regime and no longer obeys classical FDT. To extract an effective temperature, we fit the late-time part of the $\chi$ vs $C$ curve: the results of this are shown in the main text.

\begin{figure}[h!]
\begin{center}
\includegraphics[width= 0.6 \linewidth]{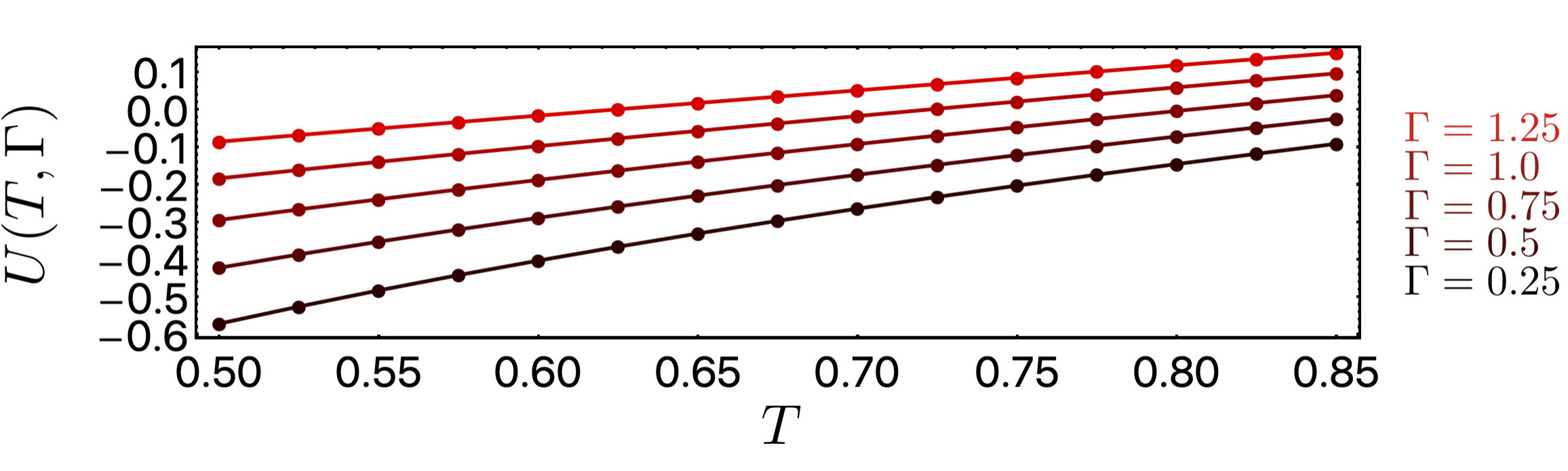}
\caption{Equilibrium internal energy of the spherical p-spin as a function of temperature $T$ and for different values of $\Gamma=\hbar^2/Jm$ controlling the strength of quantum fluctuations. We see that at fixed temperature increasing $\Gamma$ leads to an increase of the energy. Since the post-energy quench does not depend on $\hbar_F$, this immediately leads to an effective temperature decreasing with $\Gamma$.}\label{fig:U_static}
\end{center}
\end{figure}
We conclude providing further details for the thermodynamic estimate of the effective temperature discussed in the manuscript (see Figure 3 of the main text and related discussion). As discussed in the main text (see also next section), assuming thermalization after the quantum quench one can obtain the effective temperature by comparing the post-quench energy $E_Q$, which is conserved during the unitary evolution, with the internal equilibrium energy of the system $U(T,\hbar)$ at the same value of $\hbar=\hbar_F$,
\begin{equation}\label{eqn:Teff}
E_Q=U(T_{eff},\hbar_F)
\end{equation}
 We compute this internal energy within the static approximation which is valid at the relatively high temperatures under consideration~\cite{Cugliandolo+01} and is much simpler than solving the full dynamical equation in imaginary time and we plot it in Figure~\ref{fig:U_static} as a function of $\Gamma=\hbar^2/Jm$, a parameters that encodes the strength of quantum fluctuations in the spherical p-spin model\cite{Cugliandolo+01}. The key point is that increasing $\Gamma$ increases the internal energy, therefore the (effective) temperature needed to match the value of the post-quench energy $E_Q$ decreases. The numerical solution of Eq.(\ref{eqn:Teff}) leads to the \emph{thermodynamic} estimate of the effective temperature that we show in Figure 3 of the main text.

\section{Effective Temperature after a Transverse-Field Quench in the Ising p-Spin}

In this section we extend our analysis to the quantum Ising p-spin and show that our main result, namely the effective temperature decreasing with the strength of quantum fluctuation ultimately leading to enhanced glassiness, can be obtained also for this model by suddenly changing the strength of the transverse field. In order to mantain the section self-contained we briefly recall definition and basic facts about the model and then discuss how to extract the effective temperature after the quench.

\subsection{Ising p-Spin in a Transverse Field and Static Approximation}

The Hamiltonian of the Ising p-spin in a transverse field reads
\begin{equation}\label{eqn:pspin}
H(\Gamma)= -\sum_{i_1,..,i_p}J_{i_1,..,i_p}\sigma^z_{i_1}\sigma^z_{i_2}\ldots
\sigma^z_{i_p}-\Gamma\sum_i\sigma^x
\end{equation}
where $J_{i_1,..,i_p}$ are random couplings taken from a Gaussian distribution, as in Eq.~(\ref{eq.pdist}). This Hamiltonian differs from the model considered in the main text for two main features, namely (i) the degrees of freedom are true quantum spin $1/2$ thus going beyond the spherical approximation and (ii) the strength of quantum fluctuations is controlled by a well defined physical parameter in the Hamiltonian, the transverse field $\Gamma$ which can be experimentally controlled (see main text).  Those two features make the Ising p-spin substantially more challenging to solve, already in equilibrium.  Indeed although being fully connected, i.e.all spins are coupled to each others, the model in the thermodynamic limit retain a non-trivial structure. Specifically, after introducing replicas to deal with disorder average of the free energy, one can map the problem onto a set of quantum spins coupled by a self-consistent bath~\cite{Bray_1980}, similar to Dynamical Mean Field Theory~\cite{Review_DMFT_96}. Differently from the spherical case, for such a \emph{quantum impurity problem} a closed set of dynamical equations for correlation and response (in imaginary or real-time) cannot be written and one has usually to resort to approximations~\cite{GoldschmidtLaiPRL90,MillerHusePRL93} or numerical solutions~\cite{RitortAlvarez_1996,RitortPRB97,RozenbergGrempelPRL98}.  A well established one is the so called static approximation~\cite{GoldschmidtPRB90,TakahashiPRB07} which neglects the time-dependence of the bath correlation function and reduces the problem to a much simpler quantum spin in an effective static field.  Assuming a one-step replica symmetry breaking ansatz,  the free energy at inverse temperature $\beta=1/T$ within the static approximation then reads~\cite{Takahashi_2011}
\begin{align}
f = \frac{\beta J^2}{4}(p-1)(\chi^p - q^p + m q^p) - \frac{1}{\beta m} \ln \left[ \int \mathcal{D} z_1 \left( \int \mathcal{D}z_2 (\textrm{e}^{\beta h} + \textrm{e}^{-\beta h}) \right)^m \right]
\end{align}
where $q$ and $\chi$ are, respectively, the replica diagonal and off-diagonal static components of the spin-glass order parameter, $m$ is the Parisi parameter (that we will set later to zero, since we will be mainly interested in the replica symmetric phase) while we have defined
\begin{align}
h &= \sqrt{\Gamma^2 + M^2} \\
M &= J \sqrt{\frac{p}{2}} \left(\sqrt{q^{p-1}} z_1 + \sqrt{\chi^{p-1} - q^{p-1} z_2}  \right) \\
\int \mathcal{D}z_{1,2} (...) &= \int_\infty^\infty \frac{\ud z_{1,2}}{\sqrt{2 \pi}} \textrm{e}^{-z_{1,2}^2/2} (...) 
\end{align}
The equilibrium phase diagram is determined by taking derivatives of the free energy with respect to $q,\chi$ and $m$.
The Ising and spherical p-spin models share much of their equilibrium phase diagram, which still features a high-temperature paramagnetic phase and a glassy phase below a dynamical temperature $T_d(\Gamma)$~\cite{RitortIsingPspin,ObuchiEtAlJPS07,Takahashi_2011}. The transverse field $\Gamma$ controls the strength of quantum fluctuations, i.e. plays the same role as $\hbar$ in the spherical model, and when increased drive the transition from the glass to the paramagnetic phase.  We now show that also the out of equilibrium properties after a quench, as encoded in the effective temperature, are qualitatively the same between the Ising and the spherical model.

\subsection{Quantum Quenches and Effective Temperature}

We are interested in quantum quenches of $\Gamma$, i.e. we imagine to start in an equilibrium state of Eq.~(\ref{eqn:pspin}) at temperature $T_0$ with a value of the field $\Gamma=\Gamma_0$, that is from a density matrix $\rho_0\sim \exp\left(-H(\Gamma_0)/T_0\right)$ and let the system evolve for $t>0$ with Hamiltonian~(\ref{eqn:pspin})  with $\Gamma=\Gamma_F\neq \Gamma_0$, keeping the value of the random exchange constant for simplicity.  The average value of the Hamiltonian, measuring the energy of the system $E_Q$, is conserved during the evolution
\begin{equation}
\langle \Psi(t)\vert H\vert\Psi(t)\rangle=\mbox{Tr}\rho_0 e^{iHt}He^{-iHt}=\mbox{Tr}\rho_0 H\equiv E_Q
\end{equation} 
$E_Q$ is in general a function of the initial temperature $T_0$, the initial and final value of the transverse field $\Gamma_0,\Gamma_F$. More precisely it can be written in terms of expectations value over the initial state only
\begin{equation}
E_Q(T_0,\Gamma_0,\Gamma_F)=-\sum_{i_1,..,i_p}J_{i_1,..,i_p}\langle
 \sigma^z_{i_1}\sigma^z_{i_2}\ldots\sigma^z_{i_p}
\rangle_0
-\Gamma_F\sum_i\langle \sigma^x_i\rangle_0\equiv 
U(T_0,\Gamma_0)-\left(\Gamma_F-\Gamma_0\right)\sum_i\langle \sigma^x_i\rangle
\end{equation}
where we have introduced the internal energy of the Ising pspin, 
\begin{equation}
U(T,\Gamma)=\mbox{Tr}\rho(T,\Gamma)H(\Gamma)
\end{equation}
In order to discuss the thermalisation dynamics of the system after the quench we should in principle solve the full nonequilibrium dynamics of the Ising p-spin, which is however substantially more challenging than the spherical one and goes beyond the scope of the present study. However we can provide an estimate for the effective temperature based on purely energetic arguments, which we have shown in the main text for the spherical model to match perfectly the one obtained from solving the full dynamics. We can define the effective temperature $T_{\rm eff}$ as the temperature at which this conserved energy matches the equilibrium internal energy at the value of the final transverse field $\Gamma_F$, i.e.
\begin{equation}
E_Q(T_0,\Gamma_0,\Gamma_F)=U(T_{\rm eff},\Gamma_F)\,.
\end{equation}
In the following we will be mainly interested in how $T_{\rm eff}$ changes with $\Gamma_F$, at fixed values of $T_0$ and $\Gamma_0$ fixed.  Evaluating this quantity boils down to compute the internal energy of the Ising p-spin in a transverse field, a task that we accomplish in the next section within the static approximation.



\subsection{Energetics of the Ising p-spin in the Static Approximation}

Here we compute the internal energy of the Ising p-spin within the static approximation discussed before. Since we are interested in the high-temperature regime above the glass transition where the replica symmetry is unbroken we can directly set the Parisi parameter $m=0$, which simplifies the analysis. The internal energy can be obtained from the free energy by $U = \partial(\beta f)/ \partial \beta$ (including only the explicit dependence of $U$ on $\beta$~\cite{Cugliandolo+01}) which leads to:
\begin{align}
U &=  \frac{\beta J^2}{2} (p-1) (\chi^p - q^p) - \int \mathcal{D} z_1 \frac{   \left( \int{D} z_2 ( h \sinh(\beta h)) \right)}{\left( \int \mathcal{D} z_2 ( \cosh(\beta h)) \right)}
\end{align}

We can compute $\sum_i \langle \sigma_i^x \rangle$ from the free energy, which gives in the $m\to 0$ replica-symmetric limit 
\begin{align}
 \sum_i \langle \sigma_i^x \rangle &= \Gamma \int \mathcal{D}z_1 \left( \frac{\int \mathcal{D}z_2 \frac{\sinh (\beta h)}{h} }{ \int \mathcal{D} z_2 \cosh(\beta h)} \right)
 \end{align}
 Using this result we obtain for the post-quench energy
%
 \begin{align}
 E_Q(T_0,\Gamma_0,\Gamma_F) &= U(T_0,\Gamma_0) - (\Gamma_F-\Gamma_0) \sum_i \langle \sigma_i^x \rangle= \nn \\
 &=  \frac{\beta_0 J^2}{2} (p-1) (\chi^p - q^p) - \frac{\int \mathcal{D} z_1 \left( \int{D} z_2 ( h \sinh(\beta_0 h)) \right)}{ \int \mathcal{D} z_1 \left( \int \mathcal{D} z_2 ( \cosh(\beta_0 h)) \right)} - (\Gamma_F - \Gamma_0) \Gamma_0 \int \mathcal{D}z_1 \left( \frac{\int \mathcal{D}z_2 \frac{\sinh (\beta_0 h)}{h} }{ \int \mathcal{D} z_2 \cosh(\beta_0 h)} \right)
 \end{align}

At fixed $T_0$, $\Gamma_0$ and $\Gamma_F$, we can solve this numerically to obtain the $T_{eff}$, the effective temperature of the system after the quench, whose behavior we are now going to discuss. In the following we choose $J=1$ as unit of energy and fix $p=3$.

\begin{figure}[h!]
\begin{center}
\includegraphics[width= 0.6\linewidth]{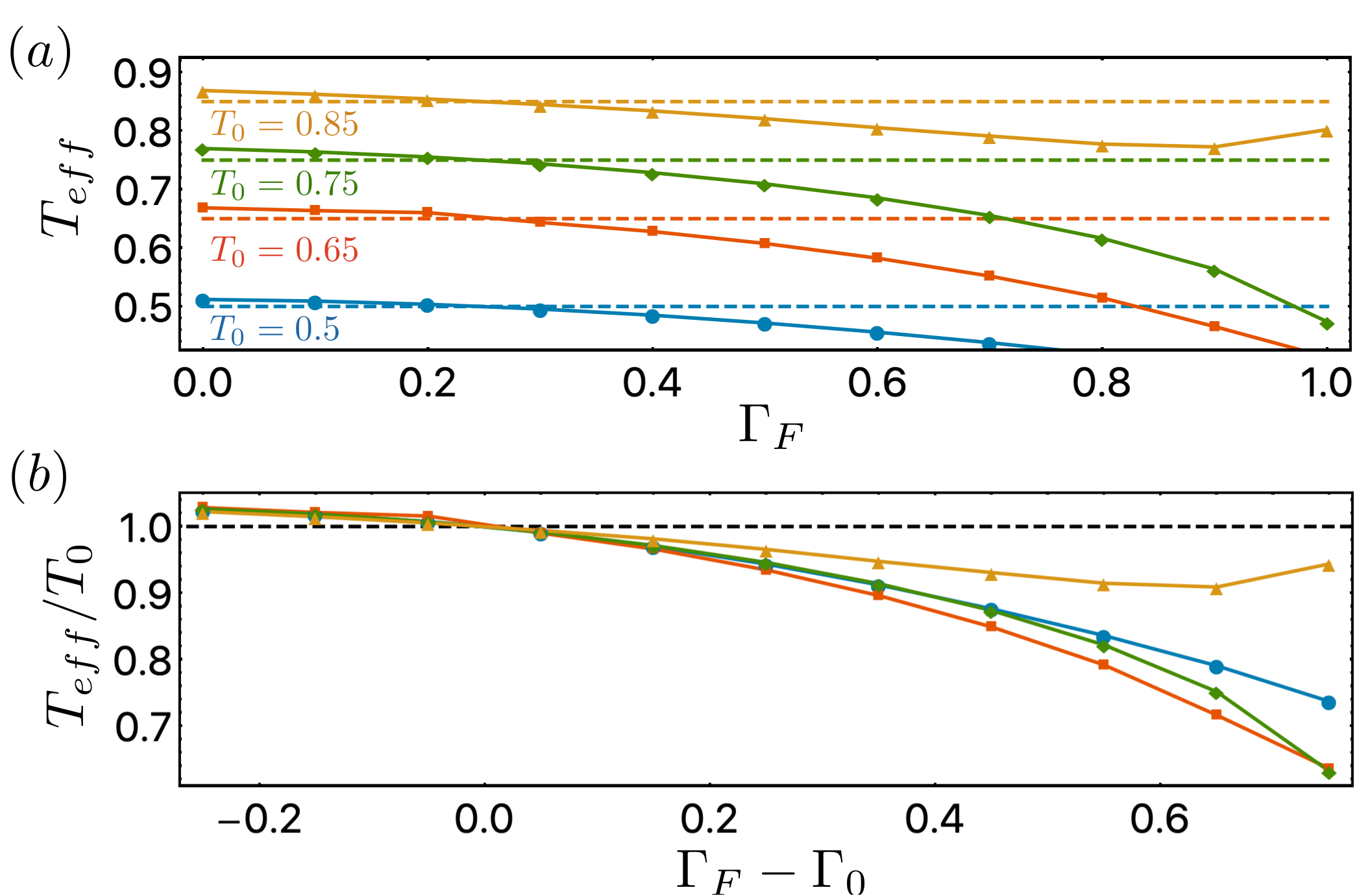}
\caption{a) Effective temperature $T_{eff}$ vs final value of transverse field $\Gamma_F$ for $\Gamma_0 = 0.25$ and for four different values of $T_0 = 0.5, 0.65, 0.75, 0.85$ (bottom to top). The dashed lines indicate the values of $T_0$ used for each color. For 
$\Gamma_F = \Gamma_0$, i.e. in absence of any quench, we find that $T_{eff}=T_0$ as expected.  b) relative temperature change, $T_{eff}/T_0$ as a function the quench amplitude, $\Gamma_F-\Gamma_0$, for different initial temperatures. From both plots we see that, quite generically, increasing the value of $\Gamma_F$ leads to a decrease of the effective temperature, i.e. cooling by quantum fluctuations, with a weak dependence on the initial temperature $T_0$. For high enough initial temperatures the effective temperature shows a non-monotonic behavior with quench amplitude.}\label{fig:Teff}
\end{center}
\end{figure}

\subsection{Results for the Effective Temperature}

In Figure~\ref{fig:Teff}a),  we plot the effective temperature $T_{eff}$ as a function of $\Gamma_F$, for $\Gamma_0=0.25$ and different initial temperatures $T_0$. We see that quite generically $T_{eff}$ is a decreasing function of $\Gamma_F$, except for high temperatures (yellow points in the plot, for $T_0=0.85$) where a non monotonous behavior appear. More importantly, if we plot the relative temperature change $T_{eff}/T_0$ versus the change in transverse field $\Gamma_F-\Gamma_0$ (bottom panel) we clearly see that the system indeed lowers its effective temperature when the transverse field is increased $\Gamma_F>\Gamma_0$ while it heats up in the opposite process. The reduced effective temperature gets closer in value  to the dynamical temperature of the Ising p-spin at which aging effects start to appear, at least based on available estimates obtained through numerical methods~\cite{Cugliandolo+04}. From this we conclude that the cooling by quantum fluctuations effect that we identified in the spherical model as key feature leading to the enhanced aging behavior consistently appears also in the Ising case.   While the dependence from the initial temperature $T_0$ is an interesting feature that deserves further investigations, for our purpose here is sufficient to show that at least in some range of temperatures $T_0$ the system cools down due to a quench of the transverse field.

%
%
%
